# Machine Learning vs. Rules and Out-of-the-Box vs. Retrained: An Evaluation of Open-Source Bibliographic Reference and Citation Parsers

### Dominika Tkaczyk
ADAPT Centre, School of Computer
Science and Statistics
Trinity College Dublin, Ireland
Dominika.Tkaczyk@adaptcentre.ie

### Andrew Collins
ADAPT Centre, School of Computer
Science and Statistics
Trinity College Dublin, Ireland
Andrew.Collins@adaptcentre.ie

### Paraic Sheridan
ADAPT Centre, School of Computer
Science and Statistics
Trinity College Dublin, Ireland
Paraic.Sheridan@adaptcentre.ie

### Joeran Beel
ADAPT Centre, School of Computer
Science and Statistics
Trinity College Dublin, Ireland
Joeran.Beel@adaptcentre.ie

## ABSTRACT

Bibliographic reference parsing refers to extracting machine-readable metadata, such as the names of the authors, the title, or journal name, from bibliographic reference strings. Many approaches to this problem have been proposed so far, including regular expressions, knowledge bases and supervised machine learning. Many open source reference parsers based on various algorithms are also available. In this paper, we apply, evaluate and compare ten reference parsing tools in a specific business use case. The tools are Anystyle-Parser, Biblio, CERMINE, Citation, Citation-Parser, GROBID, ParsCit, PDFSSA4MET, Reference Tagger and Science Parse, and we compare them in both their out-of-the-box versions and versions tuned to the project-specific data. According to our evaluation, the best performing out-of-the-box tool is GROBID (F1 0.89), followed by CERMINE (F1 0.83) and ParsCit (F1 0.75). We also found that even though machine learning-based tools and tools based on rules or regular expressions achieve on average similar precision (0.77 for ML-based tools vs. 0.76 for non-ML-based tools), applying machine learning-based tools results in a recall three times higher than in the case of non-ML-based tools (0.66 vs. 0.22). Our study also confirms that tuning the models to the task-specific data results in the increase in the quality. The retrained versions of reference parsers are in all cases better than their out-of-the-box counterparts; for GROBID F1 increased by 3% (0.92 vs. 0.89), for CERMINE by 11% (0.92 vs. 0.83), and for ParsCit by 16% (0.87 vs. 0.75).

## CCS CONCEPTS

• **Information systems** → **Information systems applications**; *Digital libraries and archives*

## KEYWORDS

bibliographic reference parsing, citation parsing, machine learning, sequence tagging

## 1 INTRODUCTION

Within the past decades there has been exponential increase in the volume of available scientific literature [1]. This has resulted in a scientific information overload problem, which refers to challenges related to consuming enormous amount of information by interested readers. Scientific information systems and digital libraries help researchers to tackle the scientific information overload problem by providing intelligent information retrieval and recommendation services. These services need machine-readable, rich bibliographic metadata of stored documents to function correctly, but this requirement is not always met in practice. As a consequence, there is a huge demand for automated methods and tools able to extract high-quality machine-readable bibliographic metadata information directly from scientific unstructured data.

Reference parsing is one important task in this research area. In reference parsing, the input is a single reference string, usually formatted in a specific bibliography style (Figure 1). The output is a machine-readable representation of the input string, typically called a parsed reference (Figure 2). Such parsed representation is a collection of metadata fields, each of which is composed of a field type (e.g. "volume" or "journal") and value (e.g. "12" or "Nature").

 

**Figure 1: An example bibliographic reference string on the input of the reference parsing algorithm.**

author: Tkaczyk, Dominika
author: Szostek, Pawel
author: Fedoryszak, Mateusz
author: Dendek, Piotr Jan
author: Bolikowski, Lukasz
title: CERMINE: automatic extraction
    of structured metadata from
    scientific literature
journal: International Journal on
    Document Analysis and Recognition
year: 2015
volume: 18
issue: 4
first page: 317
last page: 335
doi: 10.1007/s10032-015-0249-8

**Figure 2: An example output of the reference parsing task, which is a machine-readable representation of the reference string from Figure 1. This representation is a collection of metadata fields, each composed of a field type and value. For this reference, the following metadata field types were extracted: author, title, journal, volume, issue, first page, last page, year and doi.**

Bibliographic reference parsing is important for tasks such as matching citations to cited documents [2], assessing the impact of researchers [3, 4], journals [5, 6] and research institutions [7, 8], and calculating document similarity [9, 10], in the context of academic search engines [11, 12] and recommender systems [13, 14].

Reference parsing can be viewed as reversing the process of formatting a bibliography record into a string. During formatting some information is lost, and thus the reversed process is not a trivial task and usually introduces errors.

There are a few challenges related to reference parsing. First, the type of the referenced object (a journal article, a conference publication, a patent, etc.) is typically not known, so we do not know which metadata fields can be extracted. Second, the reference style is unknown, thus we do not know where in the string specific metadata fields are present. Finally, it is common for a reference string to contain errors, introduced either by humans while adding the references to the paper, or by the process of extracting the string itself from the scientific publication. These errors include for example OCR errors, unexpected spaces inside words, missing spaces, typos and errors in style-specific punctuation.

The most popular approaches to reference parsing include regular expressions, template matching, knowledge bases and supervised machine learning. There also exist a number of open source reference parsers ready to use. It is unknown, however, which approaches and which open source parsers give the best results for given metadata field types. What is more, some of the existing parsers can be tuned to the data of interest. In theory, this process should increase the quality of the results, but it is also time consuming and requires training data, which is typically expensive to obtain. An important issue is then how high an increase in the quality should be expected after retraining. These aspects are important for researchers and programmers developing larger information extraction systems for scientific data, as well as digital library practitioners wishing to use existing bibliographic reference parsers within their infrastructures.

In this study we apply, evaluate and compare a number of existing reference parsing tools, both their out-of-the-box and retrained versions, in the context of a real business project involving data from chemical domains. Specifically, we are interested in the following questions:

1. How good are reference parsing tools for our use case?
2. How do the results of machine learning-based approaches compare to the results of more static, non-trainable approaches, such as regular expressions or rules?
3. How much impact does retraining the machine learning models using project-specific data have on the parsing results?

In the following sections, we describe the state of the art, give the larger context of the business case, list the tools we evaluated, describe our evaluation setup and report the results. Finally, we discuss the findings and present conclusions.

## 2 RELATED WORK

Reference parsing is a well-known research problem, and many techniques have been proposed for solving it over the years, including regular expressions, template matching, knowledge bases and supervised machine learning.

Regular expressions are a simple way of approaching the task of reference parsing. This approach is typically based on a set of manually developed regular expressions able to capture single or multiple metadata fields in different reference styles. Such a strategy works best if the reference styles to process are known in advance and if the data contains little noise. In practice, it can be challenging to maintain a regular expressions-based system, constantly adapting the set of used regular expressions to changing data.

Regular expressions are often combined with other techniques, such as hand-crafted rules or knowledge bases. In knowledge-based approaches, at the beginning the system is populated with knowledge extracted from available data and/or existing external sources, such as digital libraries. During the actual parsing, fragments of the input reference string are matched against the information in the knowledge base. This approach works best in the case of fields which values tend to form closed sets, such as journal titles or last names.

Gupta *et al.* [15] propose a combination of regular-expression based heuristics and knowledge-based systems for reference parsing. In addition, their approach is able to match inline citations to their corresponding bibliographic references.

Constantin *et al.* [16] describe a rule- and regular expressions-based system called PDFX. PDFX is in fact a large system able to extract the logical structure of scholarly articles in PDF form, including parsed bibliography.

Day *et al.* [17] employ a hierarchical knowledge representation framework called INFOMAP for extracting metadata from reference strings. They report 92.39% accuracy for extracting author, title, journal, volume, issue, year, and page from references formatted with six major reference styles.

Finally, Cortez *et al.* [18] present FLUX-CiM, a method for reference parsing based on a knowledge base automatically constructed from an existing set of sample metadata records, obtained from public data repositories. According to their results, FLUX-CiM achieves precision and recall above 94% for a wide set of metadata fields.

In template matching approaches, references are first matched against a database of templates and then template-specific rules or regular expressions are used.

For example, Hsieh *et al.* [19] propose a reference parsing algorithm, in which the matching is based on sequence alignment. They report a 70% decrease in the average field error rate (2.24% vs. 7.54%) in comparison to a widely used machine learning-based approach.

Chen *et al.* [20] describe a tool called BibPro, which is able to extract metadata from reference strings using a gene sequence alignment tool (Basic Local Alignment Search Tool).

The most popular approach to reference parsing is supervised machine learning. In this approach training data is used to learn a so-called model, which is used during actual parsing to extract metadata from the input string. Such an approach requires little expert knowledge, as patterns are learned directly from the training data. Maintainability is also an important concern in a machine learning-based approach, however, it is comparatively easy to make sure the models are up to date by repeatedly retraining them on newer data.

In a supervised machine learning-based approach, reference parsing is usually formally defined as a sequence tagging problem. In a sequence tagging problem, on the input there is a sequence of objects represented by features, and the goal is to assign a corresponding sequence of labels, taking into account not only the features themselves, but also the dependencies between direct and indirect neighboring labels in the sequence.

For a sequence tagger to be useful for a reference parsing task, first the input reference string has to be transformed into a sequence of smaller fragments, typically called tokens. Tokenization can be performed in many different ways, for example it can be based on punctuation characters, or spaces. After tokenization, each token is assigned a label by a supervised sequence tagger. The labels usually correspond to the sought metadata field types, and a special label "other" is used for tokens that are not a part of any metadata field. Sometimes separate labels are used for the first token of a metadata field. After assigning labels to tokens, neighboring tokens with the same label are concatenated to form the final metadata fields.

It is important to note that in order to train a supervised sequence tagger for reference parsing, a specific representation of a reference string, composed of labeled tokens is required (Figure 3). In practice, training data is usually stored in an XML-based format, which can be easily transformed to the sequence of labelled tokens (Figure 4).

[AU-FN]Dominika [AU-SN]Tkaczyk [OTH], [AU-FN]Pawel [AU-SN]Szostek [OTH], [AU-FN]Mateusz [AU-SN]Fedoryszak [OTH], [AU-FN]Piotr [AU-FN]Jan [AU-SN]Dendek [OTH]and [AU-FN]Lukasz [AU-SN]Bolikowski [OTH]. [TITLE]CERMINE [TITLE]: [TITLE]automatic [TITLE]extraction [TITLE]of [TITLE]structured [TITLE]metadata [TITLE]from [TITLE]scientific [TITLE]literature [OTH]. [OTH]In [JOURNAL]International [JOURNAL]Journal [JOURNAL]on [JOURNAL]Document [JOURNAL]Analysis [JOURNAL]and [JOURNAL]Recognition [OTH], [YEAR]2015 [OTH], [OTH]vol [OTH]. [VOL]18 [OTH], [OTH]no [OTH]. [ISSUE]4 [OTH], [OTH]pp [OTH]. [FPAGE]317 [OTH]- [LPAGE]335 [OTH], [OTH]doi [OTH]: [DOI]10 [DOI]. [DOI]1007 [DOI]/ [DOI]s [DOI]10032 [DOI]- [DOI]015 [DOI]- [DOI]0249 [DOI]- [DOI]8 [OTH].

**Figure 3: An example of a reference string represented as a sequence of labelled tokens. In this case a single token can be either a sequence of letters, a sequence of digits, or a single other character. The labels are given in square brackets.**

```
<citation>
  <author><fn>Dominika</fn> <sn>Tkaczyk</sn></author>,
  <author><fn>Pawel</fn> <sn>Szostek</sn></author>,
  <author><fn>Mateusz</fn> <sn>Fedoryszak</sn></author>,
  <author><fn>Piotr Jan</fn> <sn>Dendek</sn></author> and
  <author><fn>Lukasz</fn> <sn>Bolikowski</sn></author>.
  <title> CERMINE: automatic extraction of structured metadata
          from scientific literature</title>.
  In <journal>International Journal on Document Analysis and
          Recognition</journal>,
  <year>2015</year>,
  vol. <volume>18</volume>,
  no. <issue>4</issue>,
  pp. <fpage>317</fpage>-<lpage>335</lpage>,
  doi: <doi>10.1007/s10032-015-0249-8</doi>.
</citation>
```

**Figure 4: An example of a reference string, represented in XML-based format. Given the tokenization strategy, there is a 1-1 mapping between this representation and the sequence of labelled tokens.**

Many machine learning algorithms have been applied to problem of reference parsing, including Support Vector Machines (SVM) [21, 22], Hidden Markov Models (HMM) [23, 24, 25], and Conditional Random Fields (CRF) [21, 22, 26, 27, 28, 29, 30]. SVM is a general-purpose classification technique, while both HMM and CRF can be directly employed as sequence taggers.

Hetzner [23] proposes a simple HMM-based solution for extracting metadata fields from references. Yin *et al.* [24] employ a modification of a traditional HMM called a bigram HMM, which considers words' bigram sequential relation and position information. Finally, Ojokoh *et al.* [25] explore a trigram version of HMM, reporting overall accuracy, precision, recall and F1 measure of over 95%.

By far the most popular machine learning algorithm for reference parsing is Conditional Random Fields. Councill *et al.* [26] describe ParsCit, one of the best known, widely used open source CRF-based systems for extracting metadata from references.

GROBID, created by Lopez [27], is another example of a CRF-based system able to parse bibliographic references. GROBID is also a larger tool, able to extract the metadata and logical structure from scientific papers in PDF. The author reports metadata field-level accuracy of 95,7%.

CERMINE, proposed by Tkaczyk *et al.* [28], is also a large system able to extract metadata and structure, including parsed bibliography, from scientific papers in PDF format. CERMINE's reference parsing functionality is also based on CRF technique. In 2015 CERMINE won Semantic Publishing Challenge [31, 32], which included tasks requiring accurate extraction of title and year information from bibliographic references.

Matsouka *et al.* [33] also propose a CRF-based reference parsing method, which uses lexical features as well as lexicons.

Finally, Zhang *et al.* [30] applied CRF algorithm for the task of extracting author, title, journal and year information from reference strings, reporting an overall 97.95% F1 on PubMed Central data.

Some researchers also compare various approaches to bibliographic reference parsing. For example, Zou *et al.* [21] compare CRF and SVM, achieving very similar overall accuracies for both approaches: above 99% accuracy at the token level, and over 97% accuracy at the metadata field level.

Zhang *et al.* [22] propose structural SVM with contextual features, and compare it to conventional SVM and CRF. They also report similar accuracies for all three approaches: above 98% token classification accuracy and above 95% for field extraction.

Finally, Kim *et al.* [29] describe a system called BILBO and compare it to other popular reference parsing tools (ParsCit, Biblio, free_cite and GROBID), using previously unseen data. According to their study, the best results were obtained by BILBO (F1 0.64), followed closely by GROBID (F1 0.63).

A number of reference parsers are also available as open source tools. They can be divided into two categories: tools that are solely reference parsers, and tools with wider functionality.

Pure reference parsers include:

- Anystyle-Parser [1] (a CRF-based tool written in Ruby)
- Biblio[2] (a Perl library based on regular expressions)
- BibPro[3] [20] (based on sequence alignment)
- Citation[4] (a parser written in Ruby, uses regular expressions and additional rules)
- Citation-Parser [5] (a rule-based parser written in Python)
- Free_cite[6] (a CRF-based parser written in Ruby)

- Neural Parscit[7] (a parser based on LSTM, the only deep learning-based tool we found)
- Reference Tagger[8] (a CRF-based parser written in Python)

Apart from tools providing only reference parsing functionality, there exist a few larger systems able to extract much more information from scientific documents. It is possible, however, to employ them only for the task of reference parsing. These are:

- CERMINE[9] [28]
- GROBID[10] [27]
- ParsCit[11] [26]
- PDFSSA4MET[12]
- Science Parse[13]

Table 1 summarizes the techniques employed by each parser and gives details about the extracted metadata fields.

## 3 BUSINESS CASE

Some details related to the business case are left out on purpose, as we are not allowed to publish them.

In the business project, the input is a collection of 506,540 scientific documents in PDF format, mostly from chemical domains. The goal of the project is to extract machine-readable bibliographies from the input documents in order to identify all documents cited by each paper. More specifically, for each input document we require all bibliographic items (journal papers, conference proceedings, web pages, etc.) listed in the document. Every extracted bibliographic item should be in the form of a parsed bibliographic reference.

The input documents vary in quality. Some of them are native PDF files, with all characters correctly present in the PDF content stream, while others contain the results of a separate OCR process, with typical OCR errors.

The following metadata fields are required by the client of the project as output from the reference parsing process:

- author: the first author of the referenced document, formatted as "Lastname, Initial_of_the_first_name" (e.g. "Tkaczyk, D"),
- source: the source of the referenced document, this can be the name of the journal or the conference, URL or identifier such as ArXiv id or DOI,
- year,
- volume,
- issue,
- page: the first page of the pages range,
- organization: the organization, which is an author of the referenced document, the "corporate author".

**Table 1: Summary of extracted metadata fields and techniques employed by each open source reference parser.**

| Tool | Approach | Extracted fields |
|------|----------|------------------|
| Anystyle-Parser | CRF | authors, booktitle, date, DOI, edition, editor, genre, ISBN, journal, location, pages, publisher, title, URL, volume |
| Biblio | regular expressions | authors, date, editor, genre, issue, pages, publisher, title, volume, year |
| BibPro | template matching | authors, editor, institution, issue, journal, pages, volume, year |
| CERMINE | CRF | authors, DOI, issue, pages, title, volume, year |
| citation | regular expressions + rules | authors, title, URL, year |
| Citation-Parser | rules | authors, booktitle, issue, journal, pages, publisher, title, volume, year |
| free_cite | CRF | authors, booktitle, date, editor, institution, journal, location, pages, publisher, title, volume |
| GROBID | CRF | authors, date, editor, issue, journal, organization, pages, title, volume |
| Neural ParsCit | LSTM | authors, booktitle, date, editor, institution, journal, issue, location, pages, publisher, volume |
| ParsCit | CRF | authors, booktitle, date, editor, institution, journal, issue, location, pages, publisher, volume |
| PDFSSA4MET | regular expressions | pages, title, volume, year |
| Reference Tagger | CRF | authors, issue, journal, pages, title, volume, year |
| Science Parse | CRF | author title, volume, year, journal |

Unlike the typical reference parsing task, the title of the referenced document is not required by our client. In chemistry, information about the title is often missing from the reference string, as the information about the authors, source and numbers (volume, issue, pages) are sufficient to identify a cited paper.

For the task of extracting machine-readable bibliography metadata from a scientific paper, we employ a workflow composed of three stages (Figure 5):

1. First, the PDF file is parsed and the regions containing bibliography are recognized.
2. Next, the content of these bibliography regions is split into a list of individual reference strings.
3. Finally, we perform reference parsing for each reference string separately.

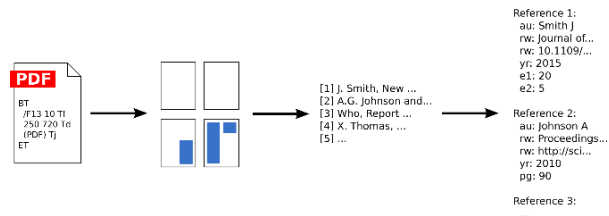

**Figure 5: The workflow of extracting machine-readable bibliography metadata from a document in PDF format. The workflow is composed of three stages: 1) recognizing the bibliography regions in the document, 2) splitting the bibliography into individual references, and 3) parsing each reference string in isolation.**

For the first two stages we employed the open source tool GROBID [27]. It uses supervised machine learning to find bibliography regions within a document and split their contents into a list of reference strings.

The third stage of the workflow is in fact reference parsing task. Since there are a lot of open source bibliographic reference parsers available (including the GROBID system itself), we decided to perform a comparative study to find out which parsers perform the best. This paper focuses solely on the third, final stage of the workflow.

## 4 METHODOLOGY

### 4.1 Evaluated Tools

In our study, we include only open source reference parsers: Anystyle-Parser, Biblio, BibPro, CERMINE, Citation, Citation-Parser, Free_cite, GROBID, Neural Parscit, ParsCit, PDFSSA4MET, Reference Tagger and Science Parse.

We were not able to evaluate three tools, due to installation errors or missing resources: BibPro, Free_cite and Neural ParsCit.

As mentioned before, not all evaluated tools extract all needed metadata fields. Also, in some cases the tools extract only a subset of a metadata field (for example, Anystyle-Parser extracts journal name, but not URL or DOI, which constitutes only part of the "source" field). Table 2 shows the matching between the fields extracted by all evaluated tools and the desired metadata fields.

**Table 2: The matching between the output of all evaluated tools and the metadata fields required in our project.**

| Tool | author | source | year | vol | issue | page | Org |
|------|--------|--------|------|-----|-------|------|-----|
| Anystyle-Parser | +[14] | +[15] | + | + | - | + | - |
| Biblio | +[16] | + | + | + | + | + | - |
| CER-MINE | + | +[17] | + | + | + | + | - |
| citation | - | +[18] | + | - | - | - | - |
| Citation-Parser | + | +[19] | + | + | + | + | - |
| GROBID | + | + | + | + | + | + | - |
| ParsCit | +[14] | +[19] | + | + | + | + | - |
| PDFSSA-4MET | - | - | + | + | - | + | - |
| Reference Tagger | + | +[15] | + | + | + | + | - |
| Science Parse | +[16] | +[15] | + | + | - | + | - |

## 4.2 Data

We had access to a collection of 9,491 pairs: PDF document + a list of parsed references, provided by our client. The collection contained 371,656 parsed references and 1,886,174 metadata fields in total. The data was manually curated and contains occasional minor errors (e.g. typos). For the purpose of the study we assume it is 100% correct.

The data was divided in the following way: roughly 67% of the dataset (6,306 documents) were used for manual exploratory analyses and training the tools, and the remaining 33% (3,185 documents) were used for testing and comparing the tools, both their out-of-the-box and retrained versions. The test set contains 64,495 references in total, which is large enough for a fair comparison.

To be useful for the evaluation and training, the data needed additional preprocessing.

For the evaluation we needed pairs: reference string + parsed reference. One problematic issue was that the client did not provide reference strings directly, but they were buried in PDF files. To obtain them, we processed the PDFs automatically using the implementation of the first two steps of our workflow (Figure 5). Unfortunately, this process is not error-free and in some cases results in strings missing or incorrect strings present on the output.

As a result, the number of extracted reference strings does not even have to be equal to the number of ground truth parsed references provided by the client, and we cannot simply use the order of the lists to decide which string corresponds to which ground truth reference. For example, the fifth reference string might correspond to the seventh parsed reference, because the first two strings are missing. To solve this problem, we used a separate process based on dynamic programming and fuzzy term matching to automatically infer correspondence between extracted strings and parsed references. This resulted in generating pairs needed for evaluation: reference string, parsed reference.

For training we needed the references in a format preserving both reference string and token tags, as explained in Section 2 (Figure 3 and Figure 4). To obtain such a representation, we matched the ground truth field values against the extracted strings, which allowed us to find substrings corresponding to the metadata fields. In some cases, this process failed to find a suitable substring (for example if the string was extracted erroneously or if it contains noise). Such references were discarded and not used for training.

## 4.3 Comparison Procedure

For a given tool and a given reference, the ground truth metadata fields are compared with the fields extracted by the tool from the string. The field values are subject to simple normalization and cleaning steps (transformation to lowercase, normalization of hyphen-like characters, cleaning fragments like "'" and "&"). After cleaning, every extracted metadata field is marked as correct or incorrect. A correct field is a field with both type and value equal to one of the fields in the ground truth parsed reference.

For a given metadata field type, we calculate precision, recall and F1 measure. Precision is the ratio of the number of correctly extracted fields (over the entire reference set) to the number of all extracted fields. Recall is the fraction of correctly extracted fields to the number of expected fields (fields in the ground truth data). F1 measure is the harmonic mean of precision and recall.

In practice, the tools vary in the field types and their meaning, and in each case careful mapping from the tool's output to our desired collection of fields was needed. For example, URL, DOI and journal name are usually present as three separate metadata field types, while in our project they are treated as one field "source". The tools also differ with respect to how the authors are extracted. Some tools (e.g. Anystyle-Parser) extract the entire author list as one field, while others split the author names. Some tools (e.g. Biblio, ParsCit, Science Parse) extract the entire author fullname as one string, while others mark additionally firstname, middlename and/or surname. In our case, the surname and first name of the first author was needed. For the systems which do not include this information, we employ additional simple heuristics on top of their output.

## 4.4 Training Procedure

Some of the tools, in particular those based on machine learning, are trainable, which means that they are able to automatically

---

[14] Entire author list only
[15] Journal name only
[16] Author fullnames only
[17] Journal name and DOI only
[18] URL only
[19] Journal name and book title only

learn custom parsing "rules" from the training data. Their out-of-the-box versions already contain trained models, which are used by default for parsing. However, we do not know whether the default models were trained on similar data as used in our project, or whether reference styles typical for chemical domains even appeared in the training sets used by the authors of the tools. As a result, we cannot be sure that the default models contain all needed information useful for parsing chemical references. For these reasons we decided to investigate whether retraining the parsers on the data specific for our project improves the parsing results.

We retrained the three most promising tools, that is, tools with the best average results obtained by their out-of-the-box versions: GROBID, CERMINE and ParsCit. For the training we used 10,000 references randomly chosen from the documents in the training set. We did not use more training data for performance reasons.

Even though all machine learning-based tools are trainable, it is important to note that they vary a lot in how easy it is to train them. For example, Anystyle-Parser, CERMINE, GROBID and ParsCit contain specific training procedures and instructions, while in other cases retraining is more difficult due to a lack of documentation.

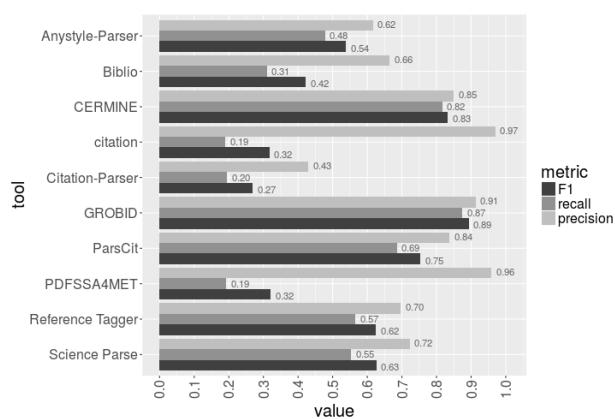

**Figure 6: The overall precision, recall and F1 values for out-of-the-box systems.**

## 5 RESULTS

Figure 6 presents the overall results of the comparison of the out-of-the-box systems and Table 3 presents the evaluation results broken down by metadata field type. Each cell in the table gives precision, recall and F1 values, respectively. For each combination (metadata type, metric) the best result is bolded. We do not give the results for organization, as none of the systems is able to extract this field.

Measured with F1, the best performing out-of-the-box tools are: GROBID (F1 0.89), followed by CERMINE (F1 0.83) and ParsCit (F1 0.75). All these tools implement CRF-based reference parsers. The worst performing systems are: Citation-Parser (F1

0.27), Citation (F1 0.32) and PDFSSA4MET (F1 0.32). All those tools are based on rules and/or regular expressions.

**Table 3: Results of the evaluation of out-of-the-box tools. Each cell gives precision, recall and F1 values, respectively. For each category the best result is bolded.**

| Tool | author | source | year | vol | issue | page |
|------|--------|--------|------|-----|-------|------|
| Anystyle-Parser | .62 | .69 | .74 | .29 | - | .76 |
| | .74 | .44 | .54 | .27 | | .69 |
| | .58 | .54 | .62 | .28 | | .73 |
| Biblio | .74 | .29 | .91 | .81 | .11 | **.99** |
| | .65 | .11 | .51 | .14 | .14 | .18 |
| | .69 | .16 | .65 | .24 | .12 | .30 |
| CERMINE | .81 | .75 | .96 | .91 | .51 | .93 |
| | .81 | .61 | .96 | .92 | **.83** | .82 |
| | .81 | .67 | .96 | .92 | .63 | .87 |
| citation | - | .47 | **.99** | - | - | - |
| | | .01 | .95 | | | |
| | | .03 | .97 | | | |
| Citation-Parser | .48 | .09 | .92 | .45 | .77 | .67 |
| | .49 | .06 | .34 | .07 | .10 | .03 |
| | .48 | .07 | .50 | .11 | .18 | .06 |
| GROBID | **.86** | **.88** | .97 | **.96** | **.91** | .90 |
| | .82 | **.82** | .95 | **.93** | .82 | **.89** |
| | .84 | **.85** | .96 | **.94** | **.86** | **.90** |
| ParsCit | **.86** | 75 | .98 | .92 | .91 | .67 |
| | **.84** | 55 | .84 | .65 | .42 | .62 |
| | **.85** | 63 | .91 | .76 | .57 | .65 |
| PDFSSA4MET | - | - | .95 | **.96** | - | **.99** |
| | | | .75 | .05 | | .20 |
| | | | .84 | .10 | | .34 |
| Reference Tagger | .69 | .63 | .97 | .51 | .58 | .76 |
| | .57 | .53 | .63 | .47 | .52 | .66 |
| | .62 | .67 | .76 | .49 | .55 | .71 |
| Science Parse | **.86** | .62 | .98 | .43 | - | .59 |
| | .62 | .52 | **.98** | .28 | | .45 |
| | .72 | .57 | **.98** | .34 | | .51 |

Measured by recall, results are the same. GROBID (0.87), CERMINE (0.82), and ParsCit (0.69) perform best. Citation (0.19), PDFSSA4MET (0.19), and Citation-Parser (0.20) perform worst.

However, measured with precision, this ranking changes. Citation (0.97), PDFSSA4MET (0.96), and GROBID (0.91) perform best, while Citation-Parser (0.43), Anystyle-Parser (0.62) and Biblio (0.66) perform worst.

In general, for all tools precision is higher than recall, with the difference ranging from 0.03 (CERMINE, 0.82 and 0.85) to 0.78

(Citation, 0.19 and 0.97). Interestingly, the difference between precision and recall is smaller in the case of machine learning-based tools (average difference 0.11) than in the case of regular expressions- or rule-based systems (average difference 0.53).

The following three systems were retrained: GROBID, CERMINE and ParsCit. These are the systems achieving the best results in the previous experiment. Figure 7 and Table 4 show the results.

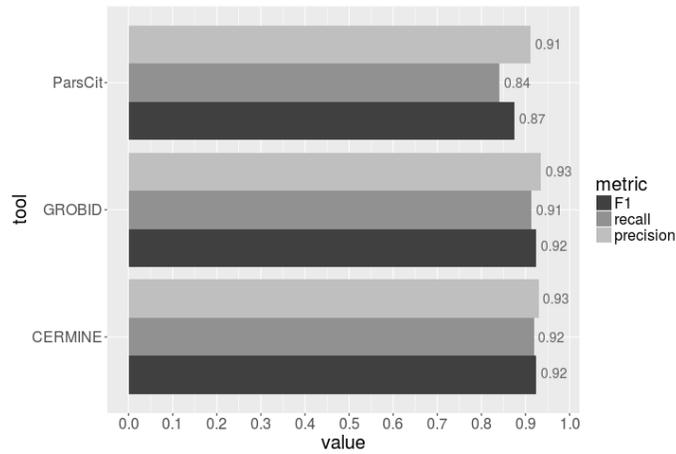

**Figure 7: The overall precision, recall and F1 of the retrained tools.**

**Table 4: Evaluation results for retrained tools, broken down by metadata field types. Each cell contains precision, recall and F1, respectively.**

| Tool | author | source | year | vol | issue | page | org |
|------|--------|--------|------|-----|-------|------|-----|
| CER-MINE | .91 | .84 | .98 | .96 | .94 | .96 | .55 |
| | .91 | .83 | .97 | .95 | .87 | .96 | .31 |
| | .91 | .84 | .98 | .96 | .90 | .96 | .39 |
| GRO-BID | .93 | .89 | .99 | .97 | .92 | .91 | .54 |
| | .92 | .85 | .98 | .95 | .87 | .89 | .52 |
| | .92 | .87 | .98 | .96 | .90 | .90 | .53 |
| Pars-Cit | .74 | .88 | .99 | .97 | .96 | .97 | - |
| | .71 | .75 | .97 | .89 | .78 | .92 | - |
| | .73 | .81 | .98 | .93 | .86 | .72 | - |

Both retrained CERMINE and GROBID achieved the same F1 of 0.92, and ParsCit was a bit worse with F1 of 0.87. The results of CERMINE and GROBID broken down by metadata types (Table 4) are similar with the exception of source (CERMINE: 0.84, GROBID: 0.87), page (CERMINE: 0.96, GROBID: 0.90) and organization (CERMINE: 0.39, GROBID: 0.53). All three systems achieved very similar high results for year. ParsCit did not extract organization at all, which suggests the training process did not pick it up from the training data.

# 6  DISCUSSION

At the beginning, we stated the following questions:

1.  How good are the results of existing reference parsing tools for previously unseen data?
2.  How do the results of machine learning-based approaches compare to the results of more static, non-trainable approaches, such as regular expressions or rules?
3.  How does retraining the machine learning models using project-specific data affect the results?

Question 1. The evaluated systems vary greatly in the quality of the results. The out-of-the-box tool achieving the best F1 is GROBID with F1 of 0.89, followed by CERMINE (F1 0.83) and ParsCit (F1 0.75). The tools with the worst F1 are: Citation-Parser (F1 0.27), Citation (F1 0.32) and PDFSSA4MET (F1 0.32). Table 5 shows the final ranking of out-of-the-box systems, ordered by decreasing F1.

**Table 5: Final ranking of out-of-the-box tools, ordered by F1.**

| Tool | F1 | precision | recall |
|------|-----|-----------|--------|
| GROBID | .89 | .91 | .87 |
| CERMINE | .83 | .85 | .82 |
| ParsCit | .75 | .84 | .69 |
| Science Parse | .63 | .72 | .55 |
| Reference Tagger | .62 | .70 | .57 |
| Anystyle-Parser | .54 | .62 | .48 |
| Biblio | .42 | .31 | .66 |
| Citation | .32 | .97 | .19 |
| PDFSSA4MET | .32 | .96 | .19 |
| Citation-Parser | .27 | .43 | .20 |

Question 2. Machine learning-based systems achieve on average better results (precision: 0.77, recall: 0.66, F1: 0.71) than regular expressions- or rule-based tools (precision: 0.76, recall: 0.22, F1: 0.33) (Figure 8). What is more, the worst ML-based tool, Anystyle-Parser (F1 0.54) outperforms the best non-ML-based tool, Biblio (F1 0.42).

The main cause of this difference is recall (Figure 8). The average recall for ML-based tools (0.66) is three times as high as non-ML-based tools (0.22). At the same time, the difference in average precisions is small (0.77 for ML-based tools vs. 0.76 for non-ML-based tools). The reason for this might be that it is relatively easy to achieve good precision of manually developed rules and regular expression, but it is difficult to have a high enough number of rules, covering all possible reference styles.

Question 3. For all three retrained systems (CERMINE, GROBID, ParsCit), retrained versions are better than out-of-the-box versions. The relative increase in F1 vary: GROBID 3% (increase from 0.89 to 0.92), CERMINE 11% (increase from 0.83 to 0.92), ParsCit 16% (increase from 0.75 to 0.87). Figure 9 compares the F1 before and after retraining for each system. In addition, in Table 6 we present the exact values for all metrics.

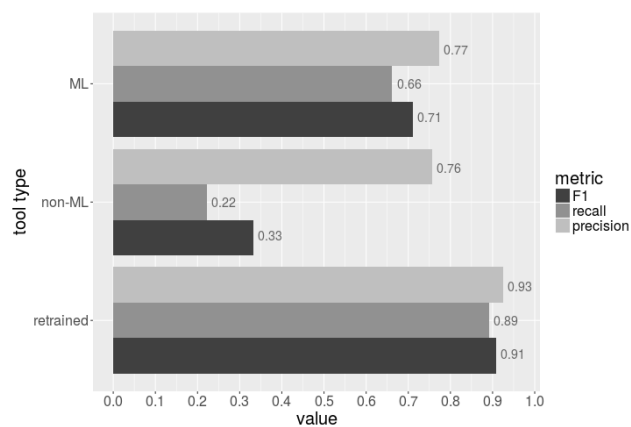

**Figure 8: Overall precision, recall and F1 aggregated by system type (ML, non-ML and retrained). The values are averaged over all tools in the respective category.**

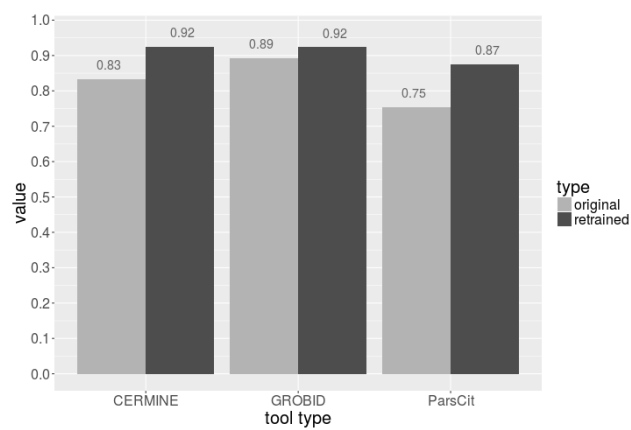

**Figure 9: Comparison of F1 for the out-of-the-box, and retrained versions of systems: CERMINE, GROBID, ParsCit.**

This effect is not surprising. We expect the retrained models to perform better, since they had an opportunity to analyze the training data and find specific "rules", as well as typical terms, useful for parsing chemical references.

We obtained the highest increase in the results in the case of ParsCit, which was the weakest system (of the three retrained) before retraining. On the other hand, in the case of GROBID the increase was the smallest. After retraining, the results of the three systems were much more similar to each other than before.

In general, our results suggest that if the pretrained version of a ML-based tool performs poorly (e.g. ParsCit), we can gain a lot by retraining the system. On the other hand, if a system already performs well (GROBID), we should still expect increase in the quality, but the magnitude of the increase might be lower.

## 7 SUMMARY AND FUTURE WORK

In this paper we study the problem of reference parsing in the context of a real business use case. We applied and compared ten reference parsing tools: Anystyle-Parser, Biblio, Citation, Citation-Parser, Reference Tagger, CERMINE, GROBID, ParsCit, PDFSSA4MET and Science Parse. We investigated the differences between tools that use rules or regular expressions and machine learning-based tools. We also checked, how important training machine learning-based tools is and how it affects the results.

**Table 6: Comparison of precision, recall and F1 between out-of-the-box versions of tools and retrained versions.**

| tool | metric | out-of-the-box | retrained | relative difference |
|---|---|---|---|---|
| GROBID | Precision | .91 | .93 | 2% |
| | Recall | .87 | .91 | 5% |
| | F1 | .89 | .92 | 3% |
| CERMINE | Precision | .85 | .93 | 9% |
| | Recall | .82 | .92 | 13% |
| | F1 | .83 | .92 | 11% |
| ParsCit | Precision | .84 | .91 | 9% |
| | Recall | .69 | .84 | 23% |
| | F1 | .75 | .87 | 16% |

According to our results, the best performing out-of-the-box tool is GROBID with F1 of 0.89, followed by CERMINE (F1 0.83) and ParsCit (F1 0.75). On average, machine learning-based systems achieve better results than rule-based systems (F1 0.71 vs. 0.33). While ML-based and non-ML-based tools achieve similar precisions (0.77 and 0.76, respectively), ML-based tools have three times higher recall than non-ML-based tools (0.66 vs. 0.22).

Our study also confirms that it is worth retraining the models using task-specific data, especially if initial results appear low. For all three retrained systems (CERMINE, GROBID, ParsCit), retrained versions are better than out-of-the-box versions, with the relative differences in F1 varying from 3% (GROBID, increase from 0.89 to 0.92), through 11% (CERMINE, increase from 0.83 to 0.92), to 16% (ParsCit, increase from 0.75 to 0.87).

It is important to note some limitations of our study. First, in our business project a very specific metadata type set was required by the client and only those types were present in the ground truth data. As a result, we did not evaluate the extraction of important metadata such as the title of the referenced document or the names of all the authors. Second, we limited our study to reference parsers fully implemented and made available on the Internet, rejecting also three systems which we could not install and use due to errors. As a result, the list of evaluated parsers does not include, for example, tools that use template matching. Finally, only the three best systems were retrained.

In the future, we plan to retrain all the available ML-based tools, and perform a similar study using other available datasets and including more metadata field types, in particular the title of the referenced paper and the names of all the authors. We also plan to experiment with building intelligent reference parsing ensembles.

## ACKNOWLEDGMENTS

This publication has emanated from research conducted with the financial support of Science Foundation Ireland (SFI) under Grant Number 13/RC/2016. The project has also received funding from the European Union's Horizon 2020 research and innovation programme under the Marie Skłodowska-Curie grant agreement No 713567.